\documentstyle[PASJadd,psfig]{PASJ95}
\draft
\begin{document}
\title{ASCA  Observations of the BL Lacertae Object OJ 287\\
in 1997 April and November}

\author{Naoki {\sc Isobe}, Makoto {\sc Tashiro}, Masahiko {\sc Sugiho},
and Kazuo {\sc Makishima}\\
{\it Department of Physics, School of Science, The  University of Tokyo,} 
{\it 7-3-1 Hongo, Bunkyo-ku, Tokyo 113-0033}\\
{\it E-mail(NI): isobe@amalthea.phys.s.u-tokyo.ac.jp}}

\abst{
X-ray properties of the BL Lacertae object OJ 287,
observed with ASCA on 1997 April 26 and November 17,
are reported.
The 0.5 -- 10 keV flux was lower than
those obtained in previous X-ray observations,
and no evidence of intensity variations
was found during each observation.
The obtained flux densities at 1 keV, $0.22 \sim 0.26 ~\mu$Jy,
exceed the extrapolations
from lower frequency synchrotron continua,
which were measured in nearly the same period
as the present ASCA observations.
The X-ray spectra acquired with the GIS and SIS were
consistently described with a single power law model
modified with the Galactic absorption,
and the derived photon indices, $1.5 \sim 1.6$,
are flatter than those observed so far.
These results strongly suggest that
the X-ray spectra observed in 1997 arise
via inverse Compton process alone.
The X-ray spectra obtained in 1994 (Idesawa et al 1997),
exhibiting a steeper slope than those in 1997,
is thought to be contaminated by "synchrotron soft tail".
}
\kword{galaxies: BL Lacertae objects: individual (OJ 287)
 ---  X-rays: general }

\maketitle
\thispagestyle{headings}

\newpage
\section{Introduction}
Blazars, including BL Lacertae objects and 
Optical Violently Variable (OVV) quasars,
are a class of active galactic nuclei 
of which the spectrum is dominated by nonthermal radiation 
from relativistic electrons in jets 
pointing at us
(e.g. Blandford, Rees 1978; Blandford, K\"{o}nigl 1979).
Their multi-frequency spectra exhibit two pronounced components.
According to the unified model of blazars,
the low energy component arises via synchrotron radiation (SR),
and the high energy one via inverse-Compton (IC) scattering
by the same population of electrons that produce the SR component.
Seed soft photons for the IC scattering can be
either the SR photons themselves
(so-called synchrotron-self-Compton process;
Maraschi, Ghisellini, Celloti 1992; Bloom, Marsher 1996),
ultraviolet photons from the accretion disk
(e.g. Dermer, Schlickheiser, Mastichiadis  1992)
or from emission line cloud
(Sikora, Begelman, Rees 1994; Blandford, Levinson 1995),
or infrared photons ambient to the host galaxy (Sikora et al. 1994).

There are two extreme subclasses of blazars
(Sambruna, Maraschi, Urry 1996; Giommi, Ansali, Micol 1995; 
 Fossati et al. 1998).
One of them, high-energy peaked BL Lac objects (HBLs; Padovani, Giommi 1996),
have the SR peak in the ultraviolet to soft X-ray band.
Their X-ray emission is dominated by the SR component
as evidence by steep spectral slopes
with photon index $\Gamma=2.0\sim3.0$ (e.g. Kubo et al. 1998).
The other, OVV quasars, show
the SR peak in the millimetor to optical range,
and their X-ray emission is dominated by
the IC component with relatively flat spectral slopes
with $\Gamma\sim1.5$ (Kubo et al. 1998).

Yet, another subclass of blazars called 
low-energy peaked BL Lac objects (LBLs; Padovani, Giommi 1996)
exhibit spectral properties intermediate
between those of HBLs and OVV quasars
(Sambruna, Maraschi, Urry 1996; Giommi, Ansali, Micol 1995; 
 Fossati et al. 1998).
In their spectra,
the SR and IC components are thought 
to compete in the ultraviolet to X-ray bands,
resulting in X-ray slopes of $\Gamma=1.5\sim2.0$ (Kubo et al. 1998).
In order to disentangle interplay between the SR and IC components,
and to examine the two component interpretation of the blazars emission,
extensive X-ray observations of LBLs have been carried out 
(e.g. Urry et al. 1996; PKS 0735+178, Madejski, Schwartz 1988;
PKS 0521-365, Garilli, Maccagni 1990; 
BL Lacertae itself, Kawai et al. 1991;
AO 0235+164, Madejski et al. 1996).
However, there are only a few LBLs of which the X-ray spectra
have been decomposed unambiguously  into the two components.

OJ 287, at a redshift of $z=0.306$ (Stickel et al. 1989),
is one of the most extensively observed BL Lac objects.
It exhibits large intensity and polarization variability 
in various energy bands,
together with superluminal motion (Gabuzda et al. 1989).
Its multi-frequency continuum is smoothly distributed 
from radio to ultraviolet bands,
with a SR peak at  $\sim 5 \times 10^{13}$ Hz
and
radio to X-ray spectral index of $\alpha_{\rm rX}\sim0.84$
(Sambruna, Maraschi, Urry 1996);
these make OJ 287 a typical LBL (Padovani, Giommi 1995).
In X-rays, 
relatively steep spectra were obtained 
with Einstein, EXOSAT and ROSAT
in 1979 -- 1980 (Madejski, Schwartz 1988),
1983 -- 1984 (Sambruna et al. 1994) and 
1991 (Comastri, Nolendi, Ghisellini 1995; Urry et al. 1996), respectively.
However, the Ginga observations performed in 1989 -- 1990 
(Ohashi 1989; Tashiro 1994)
failed to detect X-rays from this object.
The X-ray intensity varied by a factor of 3 
on time scales of months (Madejski, Schwartz 1988),
and 30\% within 3 days (Pollock et al. 1985).
The X-ray spectra obtained with the Einstein IPC,
smoothly connecting to the lower energy continua (Madejski, Schwartz 1988), 
is considered to arise from the SR process.
Giommi et al. (1995), however,
detected a hint of X-ray excess
above extrapolation from the  optical-to-ultraviolet continua.
This suggests the presence of an additional IC component
in the X-ray emission from this  LBL.

In 1994, an optical outburst of OJ 287 motivated a series of 
worldwide multi-band observations 
including an ASCA observation.
However,
the measured X-ray flux was the lowest
among the X-ray results reported so far (Idesawa et al. 1997; paper I).
Moreover, the obtained X-ray spectra, exhibiting the hardest record,
did not smoothly connect to the low energy continuum.
As argued by Idesawa et al. (paper I),
this reinforce the presence of the underlying IC component 
in the ASCA X-ray spectra,
although they could not unambiguously resolve  
the suggested IC component from the SR component.
In order to better resolve the two components,
it is important to measure long-term spectral and intensity changes
of this object,
where the two emission components are expected to vary independently
to a first approximation.
Accordingly, two separate ASCA observations were conducted in 1997.

\section{Observations and Results}
\subsection{Observation and Data Reduction}
The ASCA observations of OJ 287 were performed 
on 1997 April 26 
and  
on 1997 November 18.
The GIS
(Gas Imaging Spectrometer; Ohashi et al. 1996; Makishima et al. 1996)
and the SIS
(Solid-state-Imaging-Spectrometer; Burke et al. 1991; Gendreau et al. 1993;
Yamashita et al. 1997)
were operated in normal PH mode
and 
1CCD FAINT mode, respectively.
The target source was placed on the 1CCD nominal position of
SIS0 chip1 and SIS1 chip3.
The source was significantly detected  at the position
which coincides with the optical counterpart,
and the image was consistent with that from a point source 
within the accuracy of the ASCA point spread function.
There is no other confusing X-ray sources 
in the field of view of the GIS or SIS.
Data screening was carried out  in the standard manner,
yielding $\sim 4 \times 10^{4}$ s of exposure
from each observation.

\subsection{X-ray Light Curve}
The average 0.5--10 keV source count rates
obtained with the GIS (GIS2+GIS3) and the SIS (SIS0+SIS1),
together with statistical errors ($1\sigma$),
were
$(2.65\pm0.07)\times10^{-2}$ c s$^{-1}$ GIS$^{-1}$
and
$(4.15\pm0.09)\times10^{-2}$ c s$^{-1}$ SIS$^{-1}$
on April 26,
and 
$(3.23\pm0.07)\times10^{-2}$ c s$^{-1}$ GIS$^{-1}$
and
$(5.28\pm0.09)\times10^{-2}$ c s$^{-1}$ SIS$^{-1}$
on November 18, respectively.
These are about one third of those obtained
with the ASCA observation in 1994 (paper I).

In Figure 1, the summed GIS+SIS light curves
are shown in the 0.5--2 keV and 2--10 keV bands.
The background was not subtracted,
but is estimated to be less than
15 \% of the total count rate in each bin.
These light curves appear to exhibit intensity variations
on time scale of a few hours in each observation.
However, a chi-square test examining a constant count rate hypothesis
gives $\chi^2=21.4$ and  $6.1$  for $16$ degrees of freedom
in the hard and soft bands, respectively, for the April observation;
$\chi^2=10.9$ and $14.9$ for 15 degrees of freedom
for the November observation.
Therefore the short-term variation is concluded 
to be statistically insignificant.

\subsection{X-Ray Spectra}
In the following analysis,
the X-ray signals are  integrated within the circular regions 
centered on the source position, with the $3'$ radii for both detectors. 
The background spectra were accumulated over a source free region
of the same observation with the same radius.
Figure 2 shows the background subtracted
GIS (GIS2+GIS3) and SIS (SIS0+SIS1) spectra 
of OJ 287 
without removing the instrumental response.
Errors of these spectra represent photon statistics.
Thus, the signal X-ray were detected over the 0.6-7 keV range with the SIS,
and 0.8-10 keV  range with the GIS.

The obtained spectra are quite featureless.
A single power-law model
modified with the photoelectric absorptions
was examined against the data,
where the column density is fixed at the Galactic value,
$N_{\rm H}=2.75\times10^{20}$ cm$^{-2}$ (Elvis et al. 1989).
Then, both the GIS and SIS spectra were 
described successfully with this model,
and the derived best-fit parameters, summarized in table 1,
are consistent between the two detectors.
Accordingly, the GIS and SIS spectra were fitted simultaneously,
which was also successful (see table 1).
The derived photon indices are 
$\Gamma=1.57\pm0.09$ for April and
$\Gamma=1.51\pm0.09$ for November observations.
These values are harder than those reported
from the previous observations,
such as
$\Gamma=1.5\sim2.3$ with Einstein (Madejski, Schwartz,1988),
$\Gamma=2.16\sim2.37$ with EXOSAT (Sambruna et al. 1994),
$\Gamma=2.16\sim2.60$ with ROSAT (Comastri et al. 1995; Urry et al. 1996),
and $\Gamma=1.67\pm0.02$ with ASCA in 1994 (paper I).
The derived flux densities at 1 keV are 
$0.22\pm0.01 ~\mu$Jy for April
and
$0.25\pm0.02 ~\mu$Jy for November (see table 1).
These values are lower than those
obtained on previous occasions, such as
$0.94 \sim 2.70 ~\mu$Jy with Einstein (Madejski, Schwartz,1988),
$2.08 \sim 2.24 ~\mu$Jy with EXOSAT (Sambruna et al. 1994),
$0.44 \sim 0.9 ~\mu$Jy with ROSAT (Comastri et al. 1995; Urry et al. 1996),
and
$0.76^{+ 0.03}_{- 0.06} ~\mu$Jy with ASCA in 1994 (paper I),
all estimated at 1 keV.

\section{Discussion}
The BL Lacertae object  OJ 287 was observed twice in 1997 with ASCA, 
while the source was getting optically fainter
after the optical outburst in 1994.
The X-ray spectra obtained with the two sets of focal-plane instruments
on board ASCA have been consistently  described
with a single  power-law model 
modified with the Galactic absorption
in both observations.
The obtained source flux densities at 1 keV are lower 
and the spectra are harder 
in comparison with the previous X-ray results.
No significant evidence of short-time variability
was observed on time scales of hours.

Figure 3 shows the multi-frequency spectra of OJ 287
obtained in 1994 and 1997,
including the present X-ray results.
Radio and optical data in 1997 are presented 
by courtesy of Dr. Tapio Pursimo  
and the X-ray data in 1994 are taken from paper I.
Compared with smooth extrapolations from the lower frequency continua,
the X-ray spectra obtained with ASCA exhibit
higher fluxes and flatter slopes in both observation results.
Therefore the X-rays cannot be explained as
an extension from the lower frequency SR component.
Furthermore, the derived photon indices of $\Gamma\sim1.5$,
which are consistent with those of other well studied LBLs (Kubo et al. 1998),
are much flatter than those of the SR component in well studied HBLs,
typically $\Gamma=2.0\sim3.0$ (Sambruna et al. 1996; Kubo et al. 1998).
On the other hand,
these photon indices are closer to those of IC X-rays observed from OVV quasars
(e.g. Makino et al. 1989, 1991; Tashiro 1994; Kubo et al. 1998).
These results suggest that the X-rays from OJ 287
observed with ASCA are dominated by the IC component,
which is produced by electrons
responsible for the radio to optical SR component.

Figure 4 shows the behavior of OJ 287
on the plane of X-ray photon index versus
X-ray flux density at 1 keV ($F_{1{\rm keV}}$),
based on  available measurements taken from paper I.
Comparing the Einstein, EXOSAT, and ASCA measurements
covering similar energy bands,
the plot shows a constant photon index $\Gamma\sim1.5$
in $F_{1{\rm keV}}<0.5~\mu$Jy,
while the photon index increases with flux
for $F_{1{\rm keV}}>1~\mu$Jy.
Similar spectral behavior can be seen in some well-studied LBLs,
such as
S5 0716+714 (Giommi et al. 1999),
ON 231 (Tagliaferri et al. 2000), and
BL Lacertae (Tanihata et al. 2000), 
and is opposite to those of SR X-rays observed from HBLs
which exhibit a harder X-ray spectrum for increasing intensity.
This behavior of OJ 287 is interpreted that
a decrease in the softer  SR component unveils
the harder IC components that is less variable.
The systematic difference
between the ROSAT data points and the ASCA-Einstein-EXOSAT ones
can be explained that ROSAT, with its lower energy band,
is more sensitive to the SR component than the other experiments. 
The convergence of figure 4 at photon index $\Gamma\sim1.5$
and the stable light curve on short timescale  strongly suggest that
the X-rays observed in 1997 consist almost solely of
the IC component.

Since the X-ray spectrum observed in the 1994 optical outburst
is somewhat steeper ($\Gamma\sim1.7$; paper I)
than the limiting value of $\Gamma\sim1.5$,
it is thought to be a mixture of the SR and the IC components.
In fact, a re-analysis of the 1994 data reveals that
the spectrum above 2 keV 
is well described with a single power law model
of $\Gamma=1.54\pm0.04$ modified with the Galactic absorption
($\chi^2/{\rm d.o.f.}=209.6/232$).
This photon index is consistent with that obtained on 1997 November 18, 
when the emission is almost purely of IC origin.
However, using this parameter,
the spectrum in the full energy range of ASCA 
($0.8\sim10$ keV and $0.6\sim7.0$ keV for the GIS and SIS, respectively)
shows a soft excess below 2 keV ($\chi^2/{\rm d.o.f.}=540.0/369$).
Thus, the 1994 spectrum is examined against
a double power-law model modified with the Galactic absorption.
One power-law represents the IC component,
with photon index fixed 
at the best fit value of 1.51 in November 1997,
and normalization also fixed at the value 
determined by the single power law fit above 2 keV.
The other power-law represents the SR component,
whose spectral parameters are left free.
This model has given an acceptable fit ($\chi^2/{\rm d.o.f.}=368.1/368$)
to the 1994 ASCA spectra, yielding the parameters given in table 2.
In addition, the estimated photon index ($\Gamma \sim 2.6$)
for the SR component
is reasonable in comparison with
the observed SR components 
in HBLs (Sambruna et al. 1996; Kubo et al. 1998). 
Figure 5 shows the SR and IC components obtained in this way,
together with the multi-frequency spectrum observed in 1994. 
When  extrapolated to lower frequencies,  
the SR component smoothly connects
to the observed lower frequency SR continua.

In conclusion, as suggested by Idesawa et al. (paper I), 
the X-ray spectrum of OJ 287  is explained as a composite
between the SR and the IC components.
The present X-ray emission observed in 1997 is naturally attributed 
almost entirely to the IC emission,
while the 1994 spectrum is partly contributed by the SR component.
These results give a support to the unified scheme of blazars,
in which their multi frequency spectra are decomposed into
SR and IC components,
which compete in the X-ray band for LBLs.\par
\vspace{1pc}\par
We thank Dr. Tapio Pursimo
for providing us with the Radio/optical/ultraviolet intensity data
in advance of publication.
We also thank all the members of the ASCA team
for mission management,
including spacecraft operation,
data acquisition and instrumental calibrations.

\section*{References}
\re
Blandford R.D., Levinson A.\ 1995, ApJ.\ 441, 79
\re
Blandford R.D., Rees M.J.\ 1978 in Pittsburgh Conference on  
    	BL Lac Objects, ed A.N.\ Wolfe 
    	(University of Pittsburgh Press,Pittsburgh) p328 
\re
Blandford R.D., K$\ddot{o}$nigl A.\ 1979, ApJ.\ 232, 34
\re
Bloom S.D., Marascher A.P.\ 1996, ApJ.\ 461, 657
\re
Burke B.E., Mountain R.W., Harrison D.C. Bautz M.W.,
	Doty J.P., Ricker G.R., Daniels J.P.\ 1999, IEEE Trans.\ ED-38, 1069
\re
Comastri A., Nolendi S., Ghisellini G.\ 1995, MNRAS.\ 277, 296
\re
Dermer, C.D., Schlickeiser, R, Mastichiadis, A.\ 1992, A\&AL.\ 256, 27
\re
Elvis M., Lockman F.J., Wilkes B.J.\ 1989, AJ.\ 97, 777
\re
Fossati G., Maraschi L., Celotti A., Comastri A.,
Ghisellini G. \ 1998, MNRAS.\ 299, 433
\re
Gabuzda D.C., Wardle J.F.C., Roberts D.H.\ 1989, ApJ.\ 336, L59
\re
Garilli B., Maccagni D.\ 1990, A\&A.\ 229, 88 
\re
Gendreau K.C., Bautz M., Ricker G.\ 1993, Nucl. Instr. Meth. Phys. Res.\ A355, 318
\re
Giommi P., Ansari S.G., Micol A.\ 1995, A\&AS.\ 109, 267
\re
Giommi P., Massaro E., Chiappetti L., Ferrara E.C.
Ghisellini G., Jang M.,  Maesano M., Miller H. R.
et al.\ 1999, A\&A.\  351, 59
\re
Idesawa E., Tashiro.M, Makishima K., Kubo H.,
Otani C., Fujimoto R., KII T., Makino F. et al.\
1997, PASJ.\ 49, 631
\re
Kawai N., Matsuoka M., Bregman  J.N., Aller  H.D.,
Aller M.F., Hughes P.A., Balbus S.A., Balonek T.J.  et al.\
1991, ApJ.\ 382, 508
\re
Kubo H., Takahashi T., Madejski G.,
Tashiro M., Makino F. Inoue S. Takahara F.\
1998, ApJ.\ 504, 693
\re
Madejski G.M., Schwartz D.A.\ 1988, ApJ.\ 330, 776
\re
Madejski G.M., Takahashi T., Kubo H., Tashiro M.,
Hartman R., Kallman T., Sikora M.\ 1996,  ApJ.\ 459, 156 
\re
Makino F., Kii T., Hayashida K., Inoue H.,
Tanaka Y., Ohashi T., Makishima K., Awaki H. et al.\
1989, ApJ.\ 347, L9
\re
Makino F., Kii T., Hayashida K., Ohashi T.,
Turner M.J.L., Sadun A.C., Urry C.M., Neugebauer G. et al.\
1991 in Variability of Active Galactic Niclei,
ed H.R.\ Miller, P.J.\ Wiita
(Cambridge University Press, Cambridge) p13
\re
Makishima K., Tashiro M., Ebisawa K.,Ezawa H.,
Fukazawa Y., Gunji S., Hirayama M., Idesawa E. et al.\
1996, PASJ.\ 48, 171
\re
Maraschi L., Ghisellini G., Cellotti A.\ 1992, ApJ.\ 397, L5
\re
Ohashi T.\ 1989, in BL Lac Objects, ed L.\ Maraschi, T.\ Maccacaro,
   	 M.H.\ Ulrich (Springer-Verlag, Berlin) p297
\re
Ohashi T., Ebisawa K., Fukazawa Y., Hiyoshi K., Horii M.,
Ikebe Y., Ikeda H., Inoue H et al.\ 
1996, PASJ.\ 48, 157
\re
Padovani P., Giommi P.\ 1995, ApJ.\ 444, 567
\re
Padovani P., Giommi P.\ 1996, MNRAS.\ 279, 526
\re
Pollock A.M.T.,  Brand P.W.J.L., Bregman J.L., Robson E.I.\
1985,  Space Sci. Rev.\ 40, 607
\re
Sambruna R.M., Barr P., Giommi P., Maraschi J.L.,Robson E.I.\
1994, ApJS.\ 95, 607
\re
Sambruna R.M., Maraschi L., Urry C.M.\ 1996, ApJ.\ 463, 444
\re
Sikora M., Begelman, M.C., Rees, M.J.\ 1994, ApJ.\ 421, 153
\re
Stickel M., Fried J.W., Kuhr H.\ 1989 A\&AS.\ 80, 103
\re
Tagliaferri G., Ghisellini G., Giommi P., Chiappetti L.
Maraschi L., Celotti A., Chiaberge M., Fossati G. et al.\
2000, A\&A.\ 354, 431 
\re
Tanihata C., Takahashi T., Kataoka J., Madejski G.
Inoue S., Kubo H., Makino F., Mattox J. R., Kawai N.
2000, ApJ\ in press\ (astro-ph/0006195).
\re
Tashiro M.\ 1994, PhD Thesis, The University of Tokyo,\ ISAS RN 549
\re
Urry C.M.,Sambruna R.M., Worral D.M., Kollgaard R.I.,
Feigelson E.D., Perlman E.S., Stocke J.T.\
1996, ApJ.\ 463, 424
\re
Yamashita A., Dotani T., Bautz M., Crew G.,
Ezuka H., Gendreau K., Kotani T., Mitsuda K. et al.\
1997, IEEE Trans. Nucl. Sci.\ 44, 847

\newpage
\begin{table*}[t]
\begin{center}
Table~1. Summary of power-law fits to the X-ray spectra 
of OJ 287 obtained with ASCA in 1997.\\
\end{center}
\vspace{6pt}
\begin{tabular*}{\textwidth}{@{\hspace{\tabcolsep}
\extracolsep{\fill}}p{5pc}cccccc}
\hline\hline\\[-6pt]
                 &        \multicolumn{3}{c} {April 26}                    & \multicolumn{3}{c}{November 18 }   \\[-10pt]
                 &         \multicolumn{3}{c}{\hrulefill}                  & \multicolumn{3}{c}{\hrulefill}\\[-6pt]
Instrument       & Photon index   & $F_{\rm 1keV}~^{\ast}$         & $\chi^2/{\rm d.o.f.}$ & Photon index   & $F_{\rm 1keV}~^{\ast}$ & $\chi^2/{\rm d.o.f.}$ \\
\hline
SIS \dotfill     & $1.58\pm 0.10$ & $0.216\pm 0.014$      & 34.3/47        & $1.44\pm 0.11$ & $0.250\pm 0.017$      & 31.1/68    \\
GIS \dotfill     & $1.56\pm 0.21$ & $0.211\pm 0.035$      &  7.6/40        & $1.61\pm 0.17$ & $0.265\pm 0.032$      & 13.9/50    \\
SIS+GIS \dotfill & $1.57\pm 0.09$ & $0.215\pm 0.013$      & 42.0/89        & $1.51\pm 0.09$ & $0.252\pm 0.015$      & 47.4/120   \\
\hline
\end{tabular*}
\vspace{6pt}\par\noindent
$\ast$ Flux density at 1 keV in unit of $\mu$Jy (= $10^{-32}$ W~m$^{-2}$~Hz$^{-1}$)
\end{table*}

\begin{table*}[t]
\begin{center}
Table 2. Spectral Parameters of the IC and SR components with ASCA 
in the 1994 outburst.
\end{center}
\vspace{6pt}
\begin{tabular*}{\textwidth}{@{\hspace{\tabcolsep}
\extracolsep{\fill}}p{10pc}cc}
\hline\hline\\[-6pt]
                & Photon Index    &  $F_{\rm 1keV}~^{\ast}$ \\
\hline
 IC component \dotfill  & $1.51$ (Fixed)  &  $0.59 \pm 0.02~^{\dagger}$    \\
 SR  component \dotfill  & $2.62\pm 0.26$  &  $0.11 \pm 0.02$    \\
\hline
\end{tabular*}
\vspace{6pt}\par\noindent
$\ast$ Flux density at 1 keV in unit of $\mu$Jy (= $10^{-32}$ W~m$^{-2}$~Hz$^{-1}$).\\
$\dagger$ Determined from the single power law fit above 2 keV with photon index of 1.51. 
\end{table*}

\vspace{10cm}\
\newpage
\section*{Figure Captions}
\begin{fv}{1}{7cm}
{Summed SIS+GIS light curves for OJ 287 obtained in April
 (left) and November (right) in 1997.
 In each figure, upper panel shows the light curve in 2--10 keV and
 lower panel that in 0.5--2 keV.
 Each data point is binned into 96 min corresponding to the orbital period
 of the ASCA spacecraft.}
\end{fv}

\begin{fv}{2}{7cm}
{Background subtracted X-ray spectra of OJ 287
shown without removing the instrumental response,
observed in April (left panel) and November (right  panel)  in 1997.
Data from the two SIS detectors were summed into one spectrum
and those from the two GIS detectors were summed into the other spectrum. 
Histograms show the power-law with Galactic absorption
which best describes the GIS and SIS spectra simultaneously.
}
\end{fv}

\begin{fv}{3}{7cm}
{Multi-frequency spectra of OJ 287 obtained simultaneously 
with the ASCA observations in 1994 and 1997.}
\end{fv}

\begin{fv}{4}{7cm}
{Relations between the X-ray photon index and the X-ray flux density at 1 keV
from the available observations.
The Einstein  data are  taken from Madejski, Schwartz (1988),
EXOSAT data from Sambruna et al. (1994),
ROSAT data  from Urry et al.  (1996),
and ASCA data in 1994 from Idesawa et al. (1997).
We adopted the averaged flux density for the Einstein data.
}
\end{fv}

\begin{fv}{5}{7cm}
{The SR component and IC Component determined from the ASCA spectrum in 1994,
together with the radio and optical data in 1994 (Paper I).}
\end{fv}

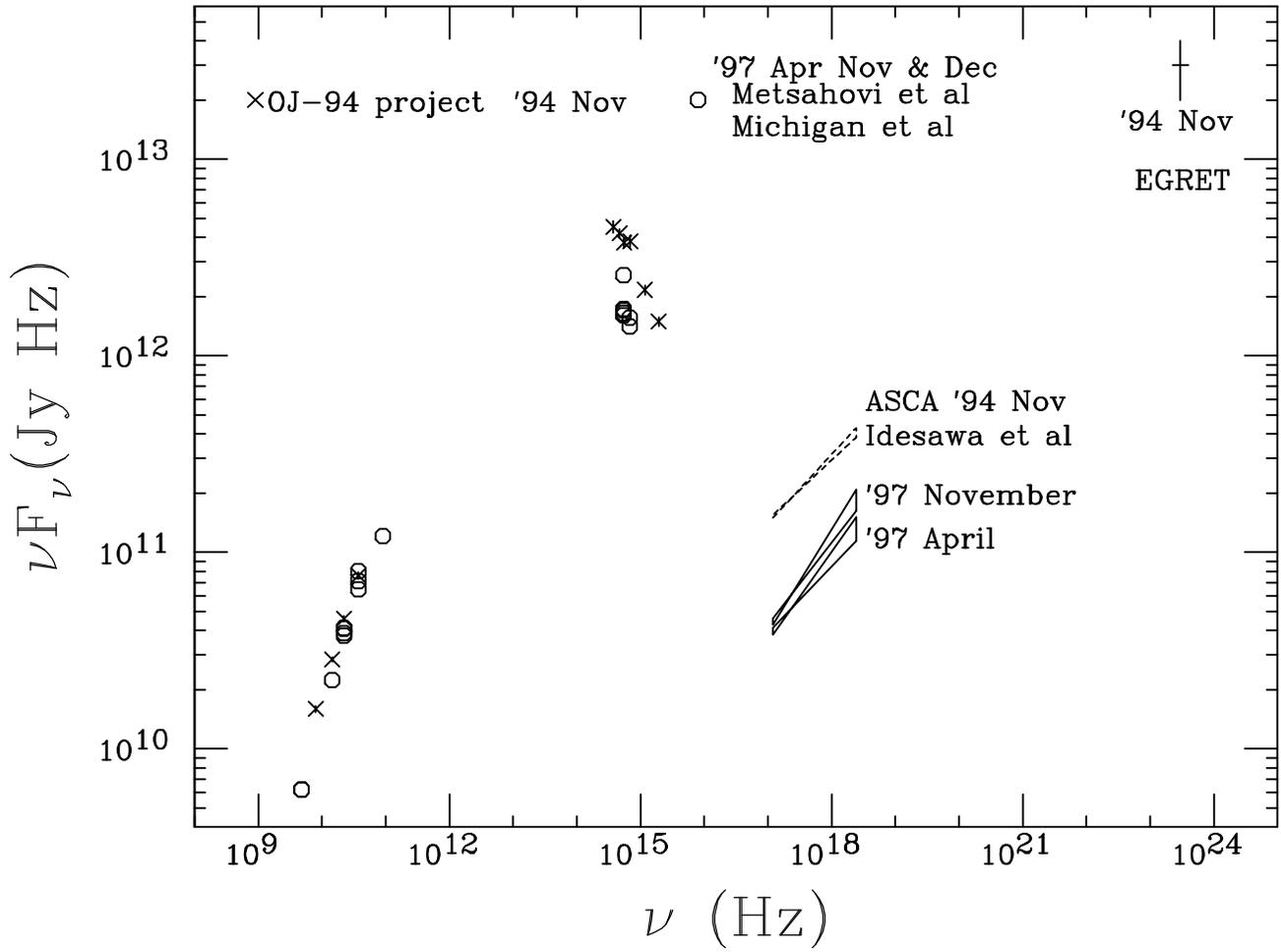
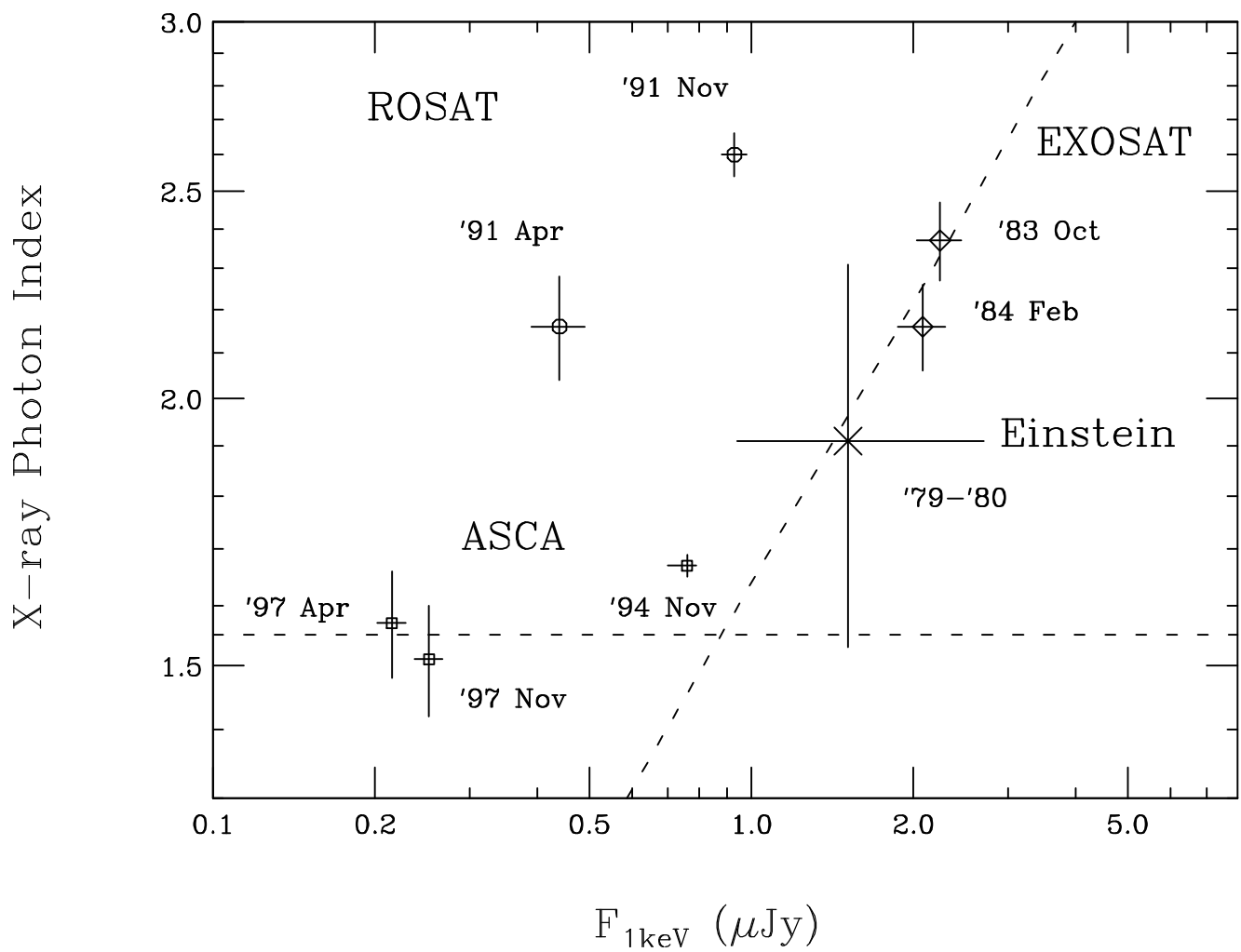
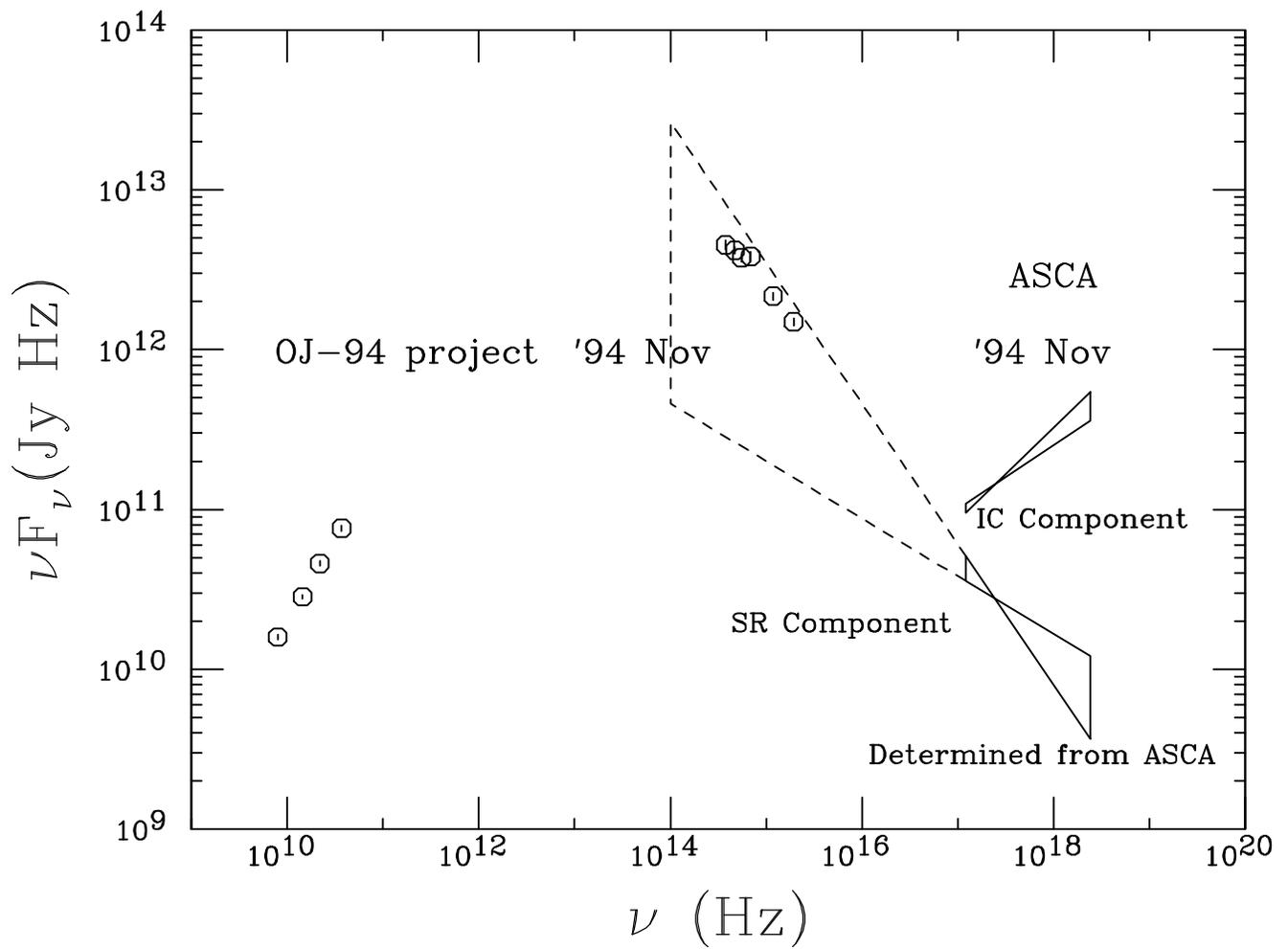
\end{document}